\documentclass[12pt,letterpaper]{article}
\usepackage{amssymb}
\usepackage{amsmath}
\usepackage{amscd}
\usepackage{graphicx}

\def\be{\begin{equation}}
\def\ee{\end{equation}}
\def\bea{\begin{eqnarray}}
\def\eea{\end{eqnarray}}
\def\bse{\begin{subequations}}
\def\ese{\end{subequations}}

\def\pa{\partial}

\def\a{{\alpha}}
\def\b{{\beta}}
\def\ad{{\dot{\alpha}}}
\def\bd{{\dot{\beta}}}
\def\D{{\rm D}}
\def\Dd{{\bar{\rm D}}}

\font\ro=cmsy10                          
\def\kcr{{\hbox{\ro \char'170}}}                
\def\ktl{{\hbox{\ro \char'170}}}        
\def\ktr{{\hbox{\ro \char'170}}}        
\def\kbl{{\hbox{\ro \char'170}}}        
\def\kbr{{\hbox{\ro \char'170}}}        

\parskip=\medskipamount                 
\thicklines                         

\newskip\humongous \humongous=0pt plus 1000pt minus 1000pt
\def\caja{\mathsurround=0pt}
\def\eqalign#1{\,\vcenter{\openup2\jot \caja
        \ialign{\strut \hfil$\displaystyle{##}$&$
        \displaystyle{{}##}$\hfil\crcr#1\crcr}}\,}
\newif\ifdtup

\def\border{                                            
        \setlength{\unitlength}{1mm}
        \newcount\xco
        \newcount\yco
        \xco=-21
        \yco=12
        \begin{picture}(140,0)
        \put(\xco,\yco){$\ktl$}
        \advance\yco by-1
        {\loop
        \put(\xco,\yco){$\kcr$}
        \advance\yco by-2
        \ifnum\yco>-240
        \repeat
        \put(\xco,\yco){$\kbl$}}
        \xco=158
        \yco=12
        \put(\xco,\yco){$\ktr$}
        \advance\yco by-1
        {\loop
        \put(\xco,\yco){$\kcr$}
        \advance\yco by-2
        \ifnum\yco>-240
        \repeat
        \put(\xco,\yco){$\kbr$}}
        \put(-20,13){\tiny **University of Maryland * Center for String and
         Particle  Theory* Physics Department***University of Maryland *Center
        for String and Particle  Theory** }
        \put(-20,-241.5){\tiny **University of Maryland * Center for String and
         Particle  Theory* Physics Department***University of Maryland *Center
        for String and Particle  Theory** }
        \end{picture}
        \par\vskip-8mm}

\def\headpic{                                           
        \indent
        \setlength{\unitlength}{.4mm}
        \thinlines
        \par
        \begin{picture}(29,16)
        \put(165,16){\line(1,0){4}}
        \put(170,16){\line(1,0){4}}
        \put(180,16){\line(1,0){4}}
        \put(175,0){\line(1,0){4}}
        \put(180,0){\line(1,0){4}}
        \put(185,0){\line(1,0){4}}
        \put(169,0){\line(0,1){16}}
        \put(170,0){\line(0,1){16}}
        \put(179,0){\line(0,1){16}}
        \put(180,0){\line(0,1){16}}
        \put(184,0){\line(0,1){16}}
        \put(185,0){\line(0,1){16}}
        \put(169,16){\oval(8,32)[bl]}
        \put(170,16){\oval(8,32)[br]}
        \put(179,0){\oval(8,32)[tl]}
        \put(185,0){\oval(8,32)[tr]}
        \end{picture}
        \par\vskip-6.5mm
        \thicklines}
\def\endtitle{\end{quotation}\newpage}                  

\begin{document}

\border\headpic {\hbox to\hsize{April 2010 \hfill
{UMDEPP 10-008}}}
\par
{$~$ \hfill
{hep-th/1004.3572}}
\par

\setlength{\oddsidemargin}{0.3in}
\setlength{\evensidemargin}{-0.3in}
\begin{center}
\vglue .10in
{\large\bf A Unified Spinorial Superfield Treatment of the\\[.1in] Higher Superspin 
Superfield Formalism \  }
\\[.5in]

S.\, James Gates, Jr.\footnote{gatess@wam.umd.edu} and
Konstantinos Koutrolikos\footnote{koutrol@umd.edu}
\\[0.2in]

{\it Center for String and Particle Theory\\
Department of Physics, University of Maryland\\
College Park, MD 20742-4111 USA}\\[2.8in]

{\bf ABSTRACT}\\[.01in]
\end{center}
\begin{quotation}
{We discuss the higher superspin superfield formalism of Kuzenko et.\ al.  from the basis of
a unified treatment of a spinorial superfield prepotential, its action and a restricted set of gauge
transformations.  We recover previous results as distinct limits of this unified treatment and 
give the first derivations of the complete set of Bianchi identities associated with these equations.}

${~~~}$ \newline
\endtitle


\section{Introduction}

~~~In 1993, two theories for higher massless superspins (one each for the cases of integer and 
half odd superspin) were developed and established in the literature \cite{Kuzenko:1993jq, Kuzenko:1993jp}.  An interesting observation is that in both theories, there exists a spinorial 
superfield (physical superfield for the integer case, auxiliary superfield for the half odd case). 
This observation motivates us to look for a theory of a spinorial superfield with some free parameters,
that in a way unifies the two previously known theories as a first step to investigate more general 
possibilities.  More specifically, for certain values of the free parameters we recover the integer 
case and for another point in the parameter space we recover the half odd case.
In the following, we will also derive the field strength superfields expressed as funcitons of
the appropriate prepotentials as well as the explicit form of their Bianchi identities. 


\section{A Proposed Action and Symmetry}
~~~The starting point would be to find the most general action for a free massless spinorial superfield $\Psi_{\a(s)\ad(s-1)}$. Furthermore since this object will be the physical superfield in the case of integer superspin, this means that the highest spin included in the supermultiplet is a fermion. Therefore the $\theta\bar{\theta}$ component of $\Psi$ has mass dimensions 3/2, which means that $\Psi$ itself has mass dimensions 1/2. For a 4D, $\cal N$ = 1 supersymmetric theory, the measure of integration over superspace $d^8z$ has mass dimensions -2, so in order our action (quadratic in $\Psi$) to be dimensionless we need 2 spinorial derivatives.

The most general action for such a spinorial superfield is
\be
\eqalign{
S=\int d^8z \Big\{&c_1\Psi^{\a(s)\ad(s-1)}\D^2\Psi_{\a(s)\ad(s-1)}+c_2\Psi^{\a(s)\ad(s-1)}\Dd^2\Psi_{\a(s)\ad(s-1)}+c.c.\cr
&\,+a_1\Psi^{\a(s)\ad(s-1)}\Dd^{\ad_s}\D_{\a_s}\bar{\Psi}_{\a(s-1)\ad(s)}+a_2\Psi^{\a(s)\ad(s-1)}\D_{\a_s}\Dd^{\ad_s}\bar{\Psi}_{\a(s-1)\ad(s)}\Big\}
}
\ee
where the parameters $a_i \in \mathbb{R}$ and $c_i \in \mathbb{C}$ in complete generality.

If this action is to describe a massless supermultiplet, it should have some gauge symmetry. We thus demand that this action is invariant under a specific gauge transformation. The way to choose this transformation is to look for symmetries that respect the highest superspin projector operator $\Pi$ \cite{Gates:1983nr}
\be
\left(\Pi\Psi\right)_{\a(2s-1)}\propto \D^{\a_{2s}}\Dd^2\D_{(\a_{2s}}\pa^{\ad_1}{}_{\a_{2s-1}}\dots\pa^{\ad_{s-1}}{}_{\a_{s+1}}\Psi_{\a(s))\ad(s-1)}
\ee

From this it is obvious there are four distinct possible gauge transformation laws of $\Psi$ that preserve the highest spin projection operator.  These individually take the forms
\bse
\bea
&&\delta\Psi_{\a(s)\ad(s-1)}=\frac{1}{s!}\D_{(\a_s}K_{\a(s-1))\ad(s-1)}\\
&&\delta\Psi_{\a(s)\ad(s-1)}=\frac{1}{(s-1)!}\Dd_{(\ad_{s-1}}\Lambda_{\a(s)\ad(s-2))}\\
&&\delta\Psi_{\a(s)\ad(s-1)}=\D^2L_{\a(s)\ad(s-1)}\\
&&\delta\Psi_{\a(s)\ad(s-1)}=\Dd^2U_{\a(s)\ad(s-1)}
\eea
\ese
and the the most general gauge transformation law can include linear combinations of these.
For the considerations of this note we pick the following gauge transformation\footnote{The general case is been discussed in  \cite{Gates:2011qa} and \cite{Gates:2011qb}}
\be
\eqalign{
\delta\Psi_{\a(s)\ad(s-1)}=\D^2L_{\a(s)\ad(s-1)}+\Dd^2U_{\a(s)\ad(s-1)}
} \label{transf.}
\ee
where $[L]=[U]=-1/2$.

Under this transformation, the different terms in the action transforms as following:
\be
\eqalign{
&\D^2\delta\Psi_{\a(s)\ad(s-1)}=\D^2\Dd^2U_{\a(s)\ad(s-1)}\cr
&\Dd^2\delta\Psi_{\a(s)\ad(s-1)}=\Dd^2\D^2L_{\a(s))\ad(s-1)}\cr
&\D^{\a_s}\Dd_{\ad_s}\delta\Psi_{\a(s)\ad(s-1)}=\D^{\a_s}\Dd_{\ad_s}\D^2L_{\a(s))\ad(s-1)}\cr
&\Dd_{\ad_s}\D^{\a_s}\delta\Psi_{\a(s)\ad(s-1)}=\Dd_{\ad_s}\D^{\a_s}\Dd^2U_{\a(s)\ad(s-1)}
~~.
}
\ee
Thus making the change of the superfield according to
\be
\Psi_{\alpha(s)\dot{\alpha}(s-1)}\rightarrow\Psi_{\alpha(s)\dot{\alpha}(s-1)}+\delta\Psi_{\alpha(s)\dot{\alpha}(s-1)}\nonumber
\ee
leads to a change in the action as:
\be
\eqalign{
\delta S =\int d^8z &{~}\Big\{\frac{2s}{s+1}c_1\Dd^{\ad_s}\D^2\Psi^{\a(s)\ad(s-1)}+a_2\D^{\a_s}\Dd^2\bar{\Psi}^{\a(s-1)\ad(s)}\Big\}\frac{1}{s!}\Dd_{(\ad_s}U_{\a(s)\ad(s-1))}\cr
&\,~+\Big\{\frac{2s}{s+1}c_1\D^{\a_s}\Dd^2\bar{\Psi}^{\a(s-1)\ad(s)}+a_2\Dd^{\ad_s}\D^2\Psi^{\a(s)\ad(s-1)}\Big\}\frac{1}{s!}\D_{(\a_s}\bar{U}_{\a(s-1))\ad(s)}\cr
&\,~+\Big\{-2c_2\D_{\a_s}\Dd^2\Psi^{\a(s)\ad(s-1)}+a_1\Dd_{\ad_s}\D^2\bar{\Psi}^{\a(s-1)\ad(s)}\Big\}\D^{\beta}L_{\beta\alpha(s-1)\dot{\alpha}(s-1)}\cr
&\,~+\Big\{-2c_2\Dd_{\ad_s}\D^2\bar{\Psi}^{\a(s-1)\ad(s)}+a_1\D_{\a_s}\Dd^2\Psi^{\a(s)\ad(s-1)}\Big\}\Dd^{\bd}\bar{L}_{\alpha(s-1)\dot{\alpha}(s-1)\bd} ~~.
}
\ee
Upon choosing (in order to construct a minimal theory) 
\bse
\bea
\frac{2s}{s+1}c_1=-a_2&&\\
2c_2=-a_1 ~~,&&
\eea
\ese
the change of the action is
\be
\eqalign{
\delta S=\int d^8z &\Big\{-a_2\Dd^{\ad_s}\D^2\Psi^{\a(s)\ad(s-1)}+a_2\D^{\a_s}\Dd^2\bar{\Psi}^{\a(s-1)\ad(s)}\Big\}\Big[\frac{1}{s!}\Dd_{(\ad_s}U_{\a(s)\ad(s-1))}\Big]\cr
&~\,+\Big\{a_2\Dd^{\ad_s}\D^2\Psi^{\a(s)\ad(s-1)}-a_2\D^{\a_s}\Dd^2\bar{\Psi}^{\a(s-1)\ad(s)}\Big\}\Big[\frac{1}{s!}\D_{(\a_s}\bar{U}_{\a(s-1))\ad(s)}\Big]\cr
&~\,+\Big\{a_1\D_{\a_s}\Dd^2\Psi^{\a(s)\ad(s-1)}+a_1\Dd_{\ad_s}\D^2\bar{\Psi}^{\a(s-1)\ad(s)}\Big\}\Big[\D^{\beta}L_{\beta\alpha(s-1)\dot{\alpha}(s-1)}\Big]\cr
&~\,+\Big\{a_1\D_{\a_s}\Dd^2\Psi^{\a(s)\ad(s-1)}+a_1\Dd_{\ad_s}\D^2\bar{\Psi}^{\a(s-1)\ad(s)}\Big\}\Big[\Dd^{\bd}\bar{L}_{\alpha(s-1)\dot{\alpha}(s-1)\bd}\Big]  ~~.
}
\ee


\section{Compensators and Bianchi Identities}
~~~To compensate for the change of the action, we introduce two real superfields $H^{(1)}_{\a(s-1)\ad(s-1)}$ and $H^{(2)}_{\a(s)\ad(s)}$
($[H^{(1)}]=[H^{(2)}]=0$) which transform:
\bea
&&\delta H^{(1)}_{\a(s-1)\ad(s-1)}=\D^{\a_s}L_{\a(s)\ad(s-1)}+\Dd^{\ad_s}\bar{L}_{\a(s-1)\ad(s)}\\
&&\delta H^{(2)}_{\a(s)\ad(s)}=\frac{1}{s!}\Dd_{(\ad_s}U_{\a(s)\ad(s-1))}-\frac{1}{s!}\D_{(\a_s}\bar{U}_{\a(s-1))\ad(s)}
\eea

and add the following terms in the action:\\
1.) a cross term (interaction between the compensators and $\Psi$)
\be
\eqalign{
S_c=\int d^8z &a_2\Big\{\Dd^{\ad_s}\D^2\Psi^{\a(s)\ad(s-1)}-\D^{\a_s}\Dd^2\bar{\Psi}^{\a(s-1)\ad(s)}\Big\}H^{(2)}_{\a(s)\ad(s)}\cr
-&a_1\Big\{\D_{\a_s}\Dd^2\Psi^{\a(s)\ad(s-1)}+\Dd_{\ad_s}\D^2\bar{\Psi}^{\a(s-1)\ad(s)}\Big\}H^{(1)}_{\a(s-1)\ad(s-1)}  ~~,
}
\ee
\noindent
2.) a kinetic energy term for $H^{(1)}$
(the most general action for a free massless superfield 
$~~~~~$ with mass dimensions zero)
\be
\eqalign{ 
{~~~~~~}
S_{k_1}=\int d^8z\Big\{
&A_1 H^{(1)}{}^{\alpha(s-1)\dot{\alpha}(s-1)}D^{\gamma}\bar{D}^2D_{\gamma}H^{(1)}_{\alpha(s-1)\dot{\alpha}(s-1)}\cr
&\,+A_2 H^{(1)}{}^{\alpha(s-1)\dot{\alpha}(s-1)}\Box H^{(1)}_{\alpha(s-1)\dot{\alpha}(s-1)}\cr
&\,+ A_3 H^{(1)}{}^{\beta\alpha(s-2)\dot{\beta}\dot{\alpha}(s-2)}[D_{\beta},\bar{D}_{\dot{\beta}}][D^{\gamma},\bar{D}^{\dot{\gamma}}]H^{(1)}_{\gamma\alpha(s-2)\dot{\gamma}\dot{\alpha}(s-2)}\cr
&\,+ A_4 H^{(1)}{}^{\beta\alpha(s-2)\dot{\beta}\dot{\alpha}(s-2)}\partial_{\beta\dot{\beta}}\partial^{\gamma\dot{\gamma}}H^{(1)}_{\gamma\alpha(s-2)\dot{\gamma}\dot{\alpha}(s-2)}\Big\}
}
\ee
$~~~~~$ and, \newline
3.) a kinetic energy term for $H^{(2)}$
\be
\eqalign{ {~~~~~~}
S_{k_2}=\int d^8z\Big\{
&B_1 H^{(2)}{}^{\alpha(s)\dot{\alpha}(s)}D^{\gamma}\bar{D}^2D_{\gamma}H^{(2)}_{\alpha(s)\dot{\alpha}(s)}\cr
&\,+B_2 H^{(2)}{}^{\alpha(s)\dot{\alpha}(s)}\Box H^{(2)}_{\alpha(s)\dot{\alpha}(s)}\cr
&\,+B_3 H^{(2)}{}^{\beta\alpha(s-1)\dot{\beta}\dot{\alpha}(s-1)}[D_{\beta},\bar{D}_{\dot{\beta}}][D^{\gamma},\bar{D}^{\dot{\gamma}}]H^{(2)}_{\gamma\alpha(s-1)\dot{\gamma}\dot{\alpha}(s-1)}\cr
&\,+B_4 H^{(2)}{}^{\beta\alpha(s-1)\dot{\beta}\dot{\alpha}(s-1)}\partial_{\beta\dot{\beta}}\partial^{\gamma\dot{\gamma}} H^{(2)}_{\gamma\alpha(s-1)\dot{\gamma}\dot{\alpha}(s-1)}\Big\} ~~~.
}
\ee
Therefore the complete action is
\be
\eqalign{
{\cal S}=&~S+S_c+S_{k_1}+S_{k_2}~~~. \cr
}
\ee

Based on this action we calculate the variations with respect each
superfield:
\be
\eqalign{
{\cal T}_{\alpha(s)\dot{\alpha}(s-1)}=&\frac{\delta {\cal S}}{\delta\Psi^{\alpha(s)\dot{\alpha}(s-1)}}\cr
=&-\frac{s+1}{s}a_2\D^2\Psi_{\a(s)\ad(s-1)}-a_1\Dd^2\Psi_{\a(s)\ad(s-1)}\cr
&+\frac{1}{s!}a_1\Dd^{\ad_s}\D_{(\a_s}\bar{\Psi}_{\a(s-1))\ad(s)}+\frac{1}{s!}a_2\D_{\a_s}\Dd^{\ad_s}\bar{\Psi}_{\a(s-1)\ad(s)}\cr
&+a_2\D^2\Dd^{\ad_s}H^{(2)}_{\a(s)\ad(s)}-\frac{1}{s!}a_1\Dd^2\D_{(\a_s}H^{(1)}_{\a(s-1))\ad(s-1)}\cr
}
\ee

\be
\eqalign{
\bar{\cal T}_{\alpha(s-1)\dot{\alpha}(s)}=&-\frac{s+1}{s}a_2\Dd^2\bar{\Psi}_{\a(s-1)\ad(s)}-a_1\D^2\bar{\Psi}_{\a(s-1)\ad(s)}\cr
&+\frac{1}{s!}a_1\D^{\a_s}\Dd_{(\ad_s}\Psi_{\a(s)\ad(s-1))}+\frac{1}{s!}a_2\Dd_{(\ad_s}\D^{\a_s}\Psi_{\a(s)\ad(s-1))}\cr
&-a_2\Dd^2\D^{\a_s}H^{(2)}_{\a(s)\ad(s)}-\frac{1}{s!}a_1\D^2\Dd_{(\ad_s}H^{(1)}_{\a(s-1)\ad(s-1))}\cr
}
\ee

\be
\eqalign{
{~~~~~~~~~}
{\cal G}^{(1)}_{\alpha(s-1)\dot{\alpha}(s-1)}=&\frac{\delta {\cal S}}{\delta H^{(1)}{}^{\alpha(s-1)\dot{\alpha}(s-1)}}\cr
=&+a_1D^{\alpha_s}\bar{D}^2\Psi_{\alpha(s)\dot{\alpha}(s-1)}+a_1\bar{D}^{\dot{\alpha}_s}D^2\bar{\Psi}_{\alpha(s-1)\dot{\alpha}(s)}\cr
&+2A_1 D^{\gamma}\bar{D}^2D_{\gamma}H^{(1)}_{\alpha(s-1)\dot{\alpha}(s-1)}+2A_2 \Box H^{(1)}_{\alpha(s-1)\dot{\alpha}(s-1)}\cr
&+\frac{2}{(s-1)!^2}A_3 [D_{(\alpha_{s-1}},\bar{D}_{(\dot{\alpha}_{s-1}}][D^{\gamma},\bar{D}^{\dot{\gamma}}]H^{(1)}_{\gamma\alpha(s-2))\dot{\gamma}\dot{\alpha}(s-2))}\cr
&+\frac{2}{(s-1)!^2}A_4 \partial_{(\alpha_{s-1}(\dot{\alpha}_{s-1}}\partial^{\gamma\dot{\gamma}} H^{(1)}_{\gamma\alpha(s-2))\dot{\gamma}\dot{\alpha}(s-2))}
}
\ee

\be
\eqalign{
{\cal G}^{(2)}_{\a(s)\ad(s)}=&\frac{\delta {\cal S}}{\delta H^{(2)}{}^{\a(s)\ad(s)}}\cr
=&+\frac{1}{s!}a_2\bar{D}_{(\dot{\alpha}_s}D^2\Psi_{\alpha(s)\dot{\alpha}(s-1))}-\frac{1}{s!}a_2D_{(\alpha_s}\bar{D}^2\bar{\Psi}_{\alpha(s-1))\dot{\alpha}(s)}\cr
&+2B_1 D^{\gamma}\bar{D}^2D_{\gamma}H^{(2)}_{\alpha(s)\dot{\alpha}(s)}+2B_2 \Box H^{(2)}_{\alpha(s)\dot{\alpha}(s)}\cr
&+\frac{2}{s!^2}B_3 [D_{(\alpha_{s}},\bar{D}_{(\dot{\alpha}_{s}}][D^{\gamma},\bar{D}^{\dot{\gamma}}]H^{(2)}_{\gamma\alpha(s-1))\dot{\gamma}\dot{\alpha}(s-1))}\cr
&+\frac{2}{s!^2}B_4 \partial_{(\alpha_{s}(\dot{\alpha}_{s}}\partial^{\gamma\dot{\gamma}} H^{(2)}_{\gamma\alpha(s-1))\dot{\gamma}\dot{\alpha}(s-1))}  ~~.
}
\ee

The gauge invariance of the action demands:
\be
\eqalign{
{~~~~} 0=\delta {\cal S}=\int d^8z \Big\{&\delta\Psi^{\alpha(s)\dot{\alpha}(s-1)}\frac{\delta {\cal S}}{\delta\Psi^{\alpha(s)\dot{\alpha}(s-1)}}
+\delta\bar{\Psi}^{\a(s-1)\ad(s)}\frac{\delta {\cal S}}{\delta\bar{\Psi}^{\alpha(s-1)\dot{\alpha}(s)}}\cr
&\,+\delta H^{(1)}{}^{\alpha(s-1)\dot{\alpha}(s-1)}\frac{\delta {\cal S}}{\delta H^{(1)}{}^{\alpha(s-1)\dot{\alpha}(s-1)}}\cr
&\,+\delta H^{(2)}{}^{\alpha(s)\dot{\alpha}(s)}\frac{\delta {\cal S}}{\delta H^{(2)}{}^{\alpha(s)\dot{\alpha}(s)}}\Big\}\cr
=\int d^8z & L^{\alpha(s)\dot{\alpha}(s-1)}\Big\{\D^2\frac{\delta {\cal S}}{\delta\Psi^{\alpha(s)\dot{\alpha}(s-1)}}-\frac{1}{s!}\D_{(\alpha_s}\frac{\delta {\cal S}}{\delta H^{(1)}{}^{\alpha(s-1))\dot{\alpha}(s-1)}}\Big\}\cr
&\,+U^{\alpha(s)\dot{\alpha}(s-1)}\Big\{\Dd^2\frac{\delta {\cal S}}{\delta\bar{\Psi}^{\alpha(s-1)\dot{\alpha}(s)}}+\Dd^{\ad_s}\frac{\delta {\cal S}}{\delta H^{(2)}{}^{\a(s)\ad(s)}}\Big\}\cr
&~+c.c.
}
\ee

Therefore the two Bianchi identities for the on-shell fields strengths ${\cal T}_{\alpha(s)\dot{\alpha}(s-1)}$,
${\cal G}^{(1)}_{\alpha(s-1)\dot{\alpha}(s-1)}$, and ${\cal G}^{(2)}_{\alpha(s)\dot{\alpha}(s)}$
are given by:
\bse
\bea
&&\D^2{\cal T}_{\alpha(s)\dot{\alpha}(s-1)}-\frac{1}{s!}\D_{(\alpha_s}{\cal G}^{(1)}_{\alpha(s-1))\dot{\alpha}(s-1)}=0\label{BE1}\\
&&\bar{\D}^2{\cal T}_{\alpha(s)\dot{\alpha}(s-1)}+\Dd^{\ad_s}{\cal G}^{(2)}_{\a(s)\ad(s)}=0\label{BE2}
\eea
\ese
and of course their complex conjugates.

The enforcement of the Bianchi identities, in other words the invariance of the full action under the gauge transformation, will fix the rest of the unconstrained parameters

Equation \eqref{BE1} gives:
\be
\eqalign{
0=&\left(2A_1-a_1\right)\frac{1}{s!}\D^2\bar{\D}^2\D_{(\alpha_s}H^{(1)}_{\alpha(s-1))\dot{\alpha}(s-1)}\cr
&-\frac{2}{s!}A_2 \Box \D_{(\alpha_s}H^{(1)}_{\alpha(s-1))\dot{\alpha}(s-1)}\cr
&-\frac{2}{s!(s-1)!}A_3 \D_{(\alpha_s}[\D_{\alpha_{s-1}},\bar{\D}_{(\dot{\alpha}_{s-1}}][\D^{\gamma},\bar{\D}^{\dot{\gamma}}]H^{(1)}_{\gamma\alpha(s-2))\dot{\gamma}\dot{\alpha}(s-2))}\cr
&-\frac{2}{s!(s-1)!}A_4 \D_{(\alpha_s}\partial_{\alpha_{s-1}(\dot{\alpha}_{s-1}}\partial^{\gamma\dot{\gamma}} H^{(1)}_{\gamma\alpha(s-2))\dot{\gamma}\dot{\alpha}(s-2))}\cr
}
\ee
which fixes the coefficients, as promised, to the following values:
\bse
\bea
&&A_1=\frac{1}{2}a_1\\
&&A_2=0\\
&&A_3=0\\
&&A_4=0 ~~.
\eea
\ese
 
 \noindent
Equation \eqref{BE2} gives:
\be
\eqalign{
0=&\left(a_2-2B_1\right)\Dd^2\D^2\Dd^{\ad_s}H^{(2)}_{\alpha(s)\dot{\alpha}(s)}\cr
&+2B_2 \Box \Dd^{\ad_s}H^{(2)}_{\alpha(s)\dot{\alpha}(s)}\cr
&+\frac{2}{s!^2}B_3 \Dd^{\ad_s}[\D_{(\alpha_{s}},\bar{\D}_{(\dot{\alpha}_{s}}][\D^{\gamma},\bar{\D}^{\dot{\gamma}}]H^{(2)}_{\gamma\alpha(s-1))\dot{\gamma}\dot{\alpha}(s-1))}\cr
&+\frac{2}{s!^2}B_4\Dd^{\ad_s} \partial_{(\alpha_{s}(\dot{\alpha}_{s}}\partial^{\gamma\dot{\gamma}} H^{(2)}_{\gamma\alpha(s-1))\dot{\gamma}\dot{\alpha}(s-1))}\cr
}
\ee
so we find that:
\bse
\bea
&&B_1=\frac{1}{2}a_2\\
&&B_2=0\\
&&B_3=0\\
&&B_4=0 ~~.
\eea
\ese

Hence the final action becomes
\be
\eqalign{
{~~~~}
S_T=&\int d^8z\Big\{-\frac{s+1}{2s}a_2\Psi^{\a(s)\ad(s-1)}\D^2\Psi_{\a(s)\ad(s-1)}+c.c.\cr
&~~~~~\,~~~~~-\frac{1}{2}a_1\Psi^{\a(s)\ad(s-1)}\Dd^2\Psi_{\a(s)\ad(s-1)}+c.c.\cr
&~~~~~\,~~~~~+a_1\Psi^{\a(s)\ad(s-1)}\Dd^{\ad_s}\D_{\a_s}\bar{\Psi}_{\a(s-1)\ad(s)}\cr
&~~~~~~~~\,~~+a_2\Psi^{\a(s)\ad(s-1)}\D_{\a_s}\Dd^{\ad_s}\bar{\Psi}_{\a(s-1)\ad(s)}\cr
&~~~~\,~~~~~~+a_2\Big(\Dd^{\ad_s}\D^2\Psi^{\a(s)\ad(s-1)}-\D^{\a_s}\Dd^2\bar{\Psi}^{\a(s-1)\ad(s)}\Big)H^{(2)}_{\a(s)\ad(s)}\cr
&~~~~~~~\,~~~-a_1\Big(\D_{\a_s}\Dd^2\Psi^{\a(s)\ad(s-1)}+\Dd_{\ad_s}\D^2\bar{\Psi}^{\a(s-1)\ad(s)}\Big)H^{(1)}_{\a(s-1)\ad(s-1)}\cr
&~~~~~~~\,~~~+\frac{1}{2}a_1 H^{(1)}{}^{\alpha(s-1)\dot{\alpha}(s-1)}D^{\gamma}\bar{D}^2D_{\gamma}H^{(1)}_{\alpha(s-1)\dot{\alpha}(s-1)}\cr
&~~~~~~\,~~~~+\frac{1}{2}a_2 H^{(2)}{}^{\alpha(s)\dot{\alpha}(s)}D^{\gamma}\bar{D}^2D_{\gamma}H^{(2)}_{\alpha(s)\dot{\alpha}(s)}\Big\}
}
\ee

and the superfield strengths take the form
\be
\eqalign{
{\cal T}_{\alpha(s)\dot{\alpha}(s-1)}=&-\frac{s+1}{s}a_2\D^2\Psi_{\a(s)\ad(s-1)}-a_1\Dd^2\Psi_{\a(s)\ad(s-1)}\cr
&+\frac{1}{s!}a_1\Dd^{\ad_s}\D_{(\a_s}\bar{\Psi}_{\a(s-1))\ad(s)}+\frac{1}{s!}a_2\D_{(\a_s}\Dd^{\ad_s}\bar{\Psi}_{\a(s-1))\ad(s)}\cr
&+a_2\D^2\Dd^{\ad_s}H^{(2)}_{\a(s)\ad(s)}-\frac{1}{s!}a_1\Dd^2\D_{(\a_s}H^{(1)}_{\a(s-1))\ad(s-1)}\cr
}
\ee

\be
\eqalign{
\bar{\cal T}_{\alpha(s-1)\dot{\alpha}(s)}=&-\frac{s+1}{s}a_2\Dd^2\bar{\Psi}_{\a(s-1)\ad(s)}-a_1\D^2\bar{\Psi}_{\a(s-1)\ad(s)}\cr
&+\frac{1}{s!}a_1\D^{\a_s}\Dd_{(\ad_s}\Psi_{\a(s)\ad(s-1))}+\frac{1}{s!}a_2\Dd_{(\ad_s}\D^{\a_s}\Psi_{\a(s)\ad(s-1))}\cr
&-a_2\Dd^2\D^{\a_s}H^{(2)}_{\a(s)\ad(s)}-\frac{1}{s!}a_1\D^2\Dd_{(\ad_s}H^{(1)}_{\a(s-1)\ad(s-1))}\cr
}
\ee

\be
\eqalign{
{\cal G}^{(1)}_{\alpha(s-1)\dot{\alpha}(s-1)}=&+a_1\D^{\alpha_s}\bar{\D}^2\Psi_{\alpha(s)\dot{\alpha}(s-1)}+a_1\bar{\D}^{\dot{\alpha}_s}\D^2\bar{\Psi}_{\alpha(s-1)\dot{\alpha}(s)}\cr
&+a_1 \D^{\gamma}\bar{\D}^2\D_{\gamma}H^{(1)}_{\alpha(s-1)\dot{\alpha}(s-1)}\cr
}
\ee

\be
\eqalign{
{\cal G}^{(2)}_{\a(s)\ad(s)}=&+\frac{1}{s!}a_2\bar{\D}_{(\dot{\alpha}_s}\D^2\Psi_{\alpha(s)\dot{\alpha}(s-1))}-\frac{1}{s!}a_2\D_{(\alpha_s}\bar{\D}^2\bar{\Psi}_{\alpha(s-1))\dot{\alpha}(s)}\cr
&+a_2 \D^{\gamma}\bar{\D}^2\D_{\gamma}H^{(2)}_{\alpha(s)\dot{\alpha}(s)}\cr
}
\ee
The form of these superfield strengths (and therefore the equations of motion on-shell) will change once we pick specific values for the parameters $a_1$ and $a_2$.


\section{Distinguished Points in the Parameter Space\\ ($a_1,a_2$)}
~~~Notice that our action has 2 unconstrained parameters $a_1,a_2$ (modulo scalings) and depending on their values the dynamics of the theory (equations of motion) change. Therefore we can say that the points in the parameter space $(a_1,a_2)$ represent different supersymmetric theories. Now we will focus on two very special points on this space, (2,0) and (0,2).  Of course, due to the possibility of re-scaling the gauge parameter superfields, these distinguished points are the only physically meaningful models described by this unified treatment.
We will prove that point (2,0) corresponds to the theory developed in  \cite{Kuzenko:1993jq} and describes a massless integer superspin supermultiplet
and the point (0,2) corresponds to the theory developed in  \cite{Kuzenko:1993jp} and describes a massless half-integer superspin supermultiplet

\subsection{Integer Superspin Action}
Consider the case where $a_1=2$ and $a_2=0$, then we recover exactly the action that appears in \cite{Kuzenko:1993jq}\footnote{with the constrained compensators expressed in terms of unconstraint prepotentials}\\
\be
\eqalign{
{\cal S}=&\int d^8z\Big\{-\Psi^{\a(s)\ad(s-1)}\Dd^2\Psi_{\a(s)\ad(s-1)}+c.c.\cr
&~~~~~~~~+2\Psi^{\a(s)\ad(s-1)}\Dd^{\ad_s}\D_{\a_s}\bar{\Psi}_{\a(s-1)\ad(s)}\cr
&~~~~~~~~-2\Big(\D_{\a_s}\Dd^2\Psi^{\a(s)\ad(s-1)}+\Dd_{\ad_s}\D^2\bar{\Psi}^{\a(s-1)\ad(s)}\Big)H^{(1)}_{\a(s-1)\ad(s-1)}\cr
&~~~~~~~~+H^{(1)}{}^{\alpha(s-1)\dot{\alpha}(s-1)}\D^{\gamma}\bar{\D}^2D_{\gamma}H^{(1)}_{\alpha(s-1)\dot{\alpha}(s-1)}\Big\}
}
\ee
By setting $a_2=0$ we make all the terms that are not invariant under the $\Dd^2U_{\a(s)\ad(s-1)}$ part of the transportation  \eqref{transf.}, vanish. That allow us to generalize a bit this piece of the transformation and still
keeping the invariance of the action.
This action is invariant under the, more general, transformations
\be
\eqalign{
&\delta\Psi_{\a(s)\ad(s-1)}=\D^2L_{\a(s)\ad(s-1)}+\frac{1}{(s-1)!}\Dd_{(\ad_{s-1}}\Lambda_{\a(s)\ad(s-2))}\cr
&\delta H^{(1)}_{\a(s-1)\ad(s-1)}=\D^{\a_s}L_{\a(s)\ad(s-1)}+\Dd^{\ad_s}\bar{L}_{\a(s-1)\ad(s)}
}
\ee
The superfield strengths take the form:
\be
\eqalign{
{\cal T}_{\alpha(s)\dot{\alpha}(s-1)}=&-2\Dd^2\Psi_{\a(s)\ad(s-1)}+\frac{2}{s!}\Dd^{\ad_s}\D_{(\a_s}\bar{\Psi}_{\a(s-1))\ad(s)}\label{T1}\cr
&-\frac{2}{s!}\Dd^2\D_{(\a_s}H^{(1)}_{\a(s-1))\ad(s-1)}\cr
}
\ee

\be
\eqalign{
\bar{\cal T}_{\alpha(s-1)\dot{\alpha}(s)}=&-2\D^2\bar{\Psi}_{\a(s-1)\ad(s)}+\frac{2}{s!}\D^{\a_s}\Dd_{(\ad_s}\Psi_{\a(s)\ad(s-1))}\label{Tb1}\cr
&-\frac{2}{s!}\D^2\Dd_{(\ad_s}H^{(1)}_{\a(s-1)\ad(s-1))}\cr
}
\ee

\be
\eqalign{
{\cal G}^{(1)}_{\alpha(s-1)\dot{\alpha}(s-1)}
=&2\D^{\alpha_s}\bar{\D}^2\Psi_{\alpha(s)\dot{\alpha}(s-1)}+2\bar{\D}^{\dot{\alpha}_s}\D^2\bar{\Psi}_{\alpha(s-1)\dot{\alpha}(s)}\label{G1}\cr
&+\frac{2}{s!}\D^{\a_s}\bar{\D}^2\D_{(\a_s}H^{(1)}_{\alpha(s-1))\dot{\alpha}(s-1)}\cr
&-2\frac{s-1}{s!}\D_{(\a_{s-1}}\Dd^2\D^{\gamma}H^{(1)}_{\gamma\alpha(s-2))\dot{\alpha}(s-1)}\cr
}
\ee

Based on \eqref{T1} and \eqref{G1} we can define:
\be
\eqalign{
{~~~~}
{\cal I}_{\a(s-1)\ad(s-1)}=&~{\cal G}^{(1)}_{\alpha(s-1)\dot{\alpha}(s-1)}+\D^{\a_s}{\cal T}_{\a(s)\ad(s-1)}\label{I1}\cr
=&-2\D^2\Dd^{\ad_s}\bar{\Psi}_{\a(s-1)\ad(s)}+2\frac{s-1}{s!}\D_{(\a_{s-1}}\Dd^{\ad_s}\D^{\gamma}\bar{\Psi}_{\gamma\a(s-2))\ad(s)}\cr
&-2\frac{s-1}{s!}\D_{(\a_{s-1}}\Dd^2\D^{\gamma}H^{(1)}_{\gamma\alpha(s-2))\dot{\alpha}(s-1)}\cr
}
\ee
and by applying to this a number of partial derivatives, contracting all the undotted indices and symmetrizing over all the dotted indices, after some algebra we find

\be
\eqalign{
&\pa^{\a_1}{}_{(\ad_{2s-2}}\dots\pa^{\a_{s-1}}{}_{\ad_s}{\cal I}_{\a(s-1)\ad(s-1))}=\label{A1}\cr
&~~~~~~~~=-\frac{2}{s}\D^2X_{\ad(2s-2)}\cr
&~~~~~~~~~~~+2\frac{s-1}{s}\pa^{\bd\b}\Dd_{(\ad_{2s-2}}\D^2\pa^{\a_1}{}_{\ad_{2s-3}}\dots\pa^{\a_{s-2}}{}_{\ad_{s}}\bar{\Psi}_{\b\a(s-2)\bd\ad(s-1))}\cr
&~~~~~~~~~~~+2i\frac{s-1}{s}\pa^{\a_1}{}_{(\ad_{2s-2}}\dots\pa^{\a_{s-2}}{}_{\ad_{s+1}}\Dd_{\ad_s}\D^2\Dd^2\D^{\b}H^{(1)}_{\b\alpha(s-2)\dot{\alpha}(s-1))}\cr
}
\ee
where
\be
\eqalign{
X_{\ad(2s-2)}=&\Dd^{\bd}\pa^{\a_1}{}_{(\ad_{2s-2}}\dots\pa^{\a_{s-1}}{}_{\ad_s}\bar{\Psi}_{\a(s-1)\bd\ad(s-1))}\cr
&+(s-1)\pa^{\bd\b}\Dd_{(\ad_{2s-2}}\pa^{\a_1}{}_{\ad_{2s-3}}\dots\pa^{\a_{s-2}}{}_{\ad_{s}}\bar{\Psi}_{\b\a(s-2)\bd\ad(s-1))  ~~. }\cr
}
\ee

\noindent
So the results in \eqref{A1} implies:
\be
\eqalign{ {~~~~}
\Dd_{(\ad_{2s-1}}\D^2X_{\ad(2s-2))}=-\frac{s}{2}(2s-2)!\Dd_{(\ad_{2s-1}}\pa^{\a_1}{}_{\ad_{2s-2}}\dots\pa^{\a_{s-1}}{}_{\ad_s}{\cal I}_{\a(s-1)\ad(s-1)) ~~.
}\label{AA1}
}
\ee
and making use of the result in \eqref{Tb1} we obtain:
\be
\eqalign{
&\Dd^2\pa^{\a_1}{}_{(\ad_{2s-1}}\dots\pa^{\a_{s-1}}{}_{\a_{s+1}}\bar{\cal T}_{\alpha(s-1)\dot{\alpha}(s))}=\label{B1}\cr
&~~~~~=-2\Dd^2\D^2\pa^{\a_1}{}_{(\ad_{2s-1}}\dots\pa^{\a_{s-1}}{}_{\a_{s+1}}\bar{\Psi}_{\a(s-1)\ad(s))}\cr
&~~~~~~~+2i\pa^{\a_1}{}_{(\ad_{2s-1}}\dots\pa^{\a_{s-1}}{}_{\a_{s+1}}\pa^{\a_s}{}_{\ad_{s}}\Psi_{\a(s)\ad(s-1))}\cr
&~~~~~~~-2\Dd^2\D^2\Dd_{(\ad_{2s-1}}\pa^{\a_1}{}_{\ad_{2s-2}}\dots\pa^{\a_{s-1}}{}_{\a_s}H^{(1)}_{\a(s-1)\ad(s-1))} ~~.
}
\ee

Finally  \eqref{T1} yields:
\be
\eqalign{
&\pa^{\a_1}{}_{(\ad_{2s-1}}\dots\pa^{\a_{s}}{}_{\ad_{s}}{\cal T}_{\a(s)\ad(s-1))}=\label{C1}\cr
&~~~~=-2\Dd^2\pa^{\a_1}{}_{(\ad_{2s-1}}\dots\pa^{\a_{s}}{}_{\ad_{s}}\Psi_{\a(s)\ad(s-1))}\cr
&~~~~~~+2i\Dd^{\bd}\D^2\Dd_{(\ad_{2s-1}}\pa^{\a_1}{}_{\ad_{2s-2}}\dots\pa^{\a_{s-1}}{}_{\ad_{s}}\bar{\Psi}_{\a(s-1)\ad(s-1))\bd}\cr
&~~~~~~-2i\Dd^2\D^2\pa^{\a_1}{}_{(\ad_{2s-1}}\dots\pa^{\a_{s-1}}{}_{\ad_{s+1}}\bar{\Psi}_{\a(s-1)\ad(s))}\cr
&~~~~~~-2i\Dd^2\D^2\Dd_{(\ad_{2s-1}}\pa^{\a_1}{}_{\ad_{2s-2}}\dots\pa^{\a_{s-1}}{}_{\ad_{s}}H^{(1)}_{\a(s-1)\ad(s-1))}\cr
}
\ee

Based on \eqref{B1} ,the identity:
\be
\eqalign{
{~~~~~}
\Dd_{(\ad_{2s-1}}\pa^{\a_1}{}_{\ad_{2s-2}}\dots\pa^{\a_{s-1}}{}_{\ad_{s}}&\bar{\Psi}_{\a(s-1)\ad(s-1))\bd}~=\cr
&~\left[\frac{1}{2s}\right]\Dd_{(\ad_{2s-1}}\pa^{\a_1}{}_{\ad_{2s-2}}\dots
\pa^{\a_{s-1}}{}_{\ad_{s}}\bar{\Psi}_{\a(s-1)\ad(s-1)\bd)}\cr
&~+\left[\frac{2s-1}{(2s)!}\right]C_{\bd(\ad_{2s-1}}X_{\ad(2s-2))}
}
\ee
and the definition
\be
\eqalign{
\bar{\cal W}_{\ad(2s)}=\D^2\Dd_{(\ad_{2s}}\pa^{\a_1}{}_{\ad_{2s-1}}\dots\pa^{\a_{s-1}}{}_{\ad_{s+1}}\bar{\Psi}_{\a(s-1)\ad(s))} ~~,
}
\ee
 equation \eqref{C1} becomes
  \be
 \eqalign{
 &\pa^{\a_1}{}_{\ad_{2s-1}}\dots\pa^{\a_{s}}{}_{\ad_{s}}{\cal T}_{\a(s)\ad(s-1))}=\label{D1}\cr
 &~~~~=i\Dd^2\pa^{\a_1}{}_{(\ad_{2s-1}}\dots\pa^{\a_{s-1}}{}_{\a_{s+1}}\bar{\cal T}_{\alpha(s-1)\dot{\alpha}(s))}\cr
 &~~~~~+2i\frac{2s-1}{(2s)!}\Dd_{(\ad_{2s-1}}D^2X_{\ad_{2s-2})}\cr
 &~~~~~+\frac{i}{s}\Dd^{\ad_{2s}}\bar{\cal W}_{\ad(2s)}
 }
 \ee
 so that with the help of  \eqref{AA1} we obtain the relation between the physical field strength
 superfield $\bar{\cal W}_{\ad(2s)}$ and the on-shell field strength superfields:
 \be
 \eqalign{
 \Dd^{\ad_{2s}}\bar{\cal W}_{\ad(2s)}
 =&\frac{s}{2}\Dd_{(\ad_{2s-1}}\pa^{\a_1}{}_{\ad_{2s-2}}\dots\pa^{\a_{s-1}}{}_{\ad_s}\D^{\b}{\cal T}_{\b\a(s-1)\ad(s-1))}\cr
 &-is\pa^{\a_1}{}_{\ad_{2s-1}}\dots\pa^{\a_{s}}{}_{\ad_{s}}{\cal T}_{\a(s)\ad(s-1))}\cr
 &-s\Dd^2\pa^{\a_1}{}_{(\ad_{2s-1}}\dots\pa^{\a_{s-1}}{}_{\a_{s+1}}\bar{\cal T}_{\alpha(s-1)\dot{\alpha}(s))}\cr
 &\frac{s}{2}\Dd_{(\ad_{2s-1}}\pa^{\a_1}{}_{\ad_{2s-2}}\dots\pa^{\a_{s-1}}{}_{\ad_s}{\cal G}^{(1)}_{\a(s-1)\ad(s-1)} ~~~.
}
 \ee

Also by its definition we find that:
\be
\eqalign{
\D_{\a}\bar{\cal W}_{\ad(2s)}=0
}
\ee
Appropriately on-shell the field strength superfields indicated immediately below
\bse
\bea
&&{\cal T}_{\a(s)\ad(s-1)}=0\\
&&{\cal G}^{(1)}_{\a(s-1)\ad(s-1)}=0
\eea
\ese
vanish.  Hence we find that
\bse
\bea
&\Dd^{\ad_{2s}}\bar{\cal W}_{\ad(2s)}=0 ~,~ &\D^{\a_{2s}}{\cal W}_{\a(2s)}=0\\
&\D_{\a}\bar{\cal W}_{\ad(2s)}=0 ~,~ &\Dd_{\ad}{\cal W}_{\a(2s)}=0
\eea
\ese
These are exactly the equations of motion needed \cite{Gates:1983nr, Ogievetsky:1976qb, Sokatchev:1975gg} in order to describe an integer superspin $Y=s$ massless supermultiplet

\subsection{Half-Odd Superspin Action}

The case where $a_1=0$ and $a_2=2$, gives back the action proposed in \cite{Kuzenko:1993jp}
\be
\eqalign{
{\cal S}=&\int d^8z\Big\{-\frac{s+1}{s}\Psi^{\a(s)\ad(s-1)}\D^2\Psi_{\a(s)\ad(s-1)}+c.c.\cr
&~~~~~~~~+2\Psi^{\a(s)\ad(s-1)}\D_{\a_s}\Dd^{\ad_s}\bar{\Psi}_{\a(s-1)\ad(s)}\cr
&~~~~~~~~+2\Big(\Dd^{\ad_s}\D^2\Psi^{\a(s)\ad(s-1)}-\D^{\a_s}\Dd^2\bar{\Psi}^{\a(s-1)\ad(s)}\Big)H^{(2)}_{\a(s)\ad(s)}\cr
&~~~~~~~~+H^{(2)}{}^{\alpha(s)\dot{\alpha}(s)}\D^{\gamma}\bar{\D}^2\D_{\gamma}H^{(2)}_{\alpha(s)\dot{\alpha}(s)}\Big\}
}
\ee
By setting $a_1=0$, we make the terms which are not invariant under the $\D^2 L_{\a(s)\ad(s-1)}$ piece of \eqref{transf.} to vanish. So this piece of the transformation can be generalized without losing the invariance of the action. This action is invariant under the following transformations
\be
\eqalign{
&\delta H^{(2)}_{\a(s)\ad(s)}=\frac{1}{s!}\Dd_{(\ad_s}U_{\a(s)\ad(s-1))}-\frac{1}{s!}\D_{(\a_s}
\bar{U}_{\a(s-1))\ad(s)}\cr
&\delta\Psi_{\a(s)\ad(s-1)}=\Dd^2U_{\a(s)\ad(s-1)}+\D^{\a_{s+1}}\Lambda_{\a(s+1)\ad(s)}
}
\ee
The superfield strengths are:
\be
\eqalign{
{\cal T}_{\alpha(s)\dot{\alpha}(s-1)}=&-2\frac{s+1}{s}\D^2\Psi_{\a(s)\ad(s-1)}+2\frac{1}{s!}\D_{(\a_s}\Dd^{\ad_s}\bar{\Psi}_{\a(s-1))\ad(s)}\label{T2}\cr
&+2\D^2\Dd^{\ad_s}H^{(2)}_{\a(s)\ad(s)}\cr
}
\ee

\be
\eqalign{
\bar{\cal T}_{\alpha(s-1)\dot{\alpha}(s)}=&-2\frac{s+1}{s}\Dd^2\bar{\Psi}_{\a(s-1)\ad(s)}+\frac{2}{s!}\Dd_{(\ad_s}\D^{\a_s}\Psi_{\a(s)\ad(s-1))}\label{Tb2}\cr
&-2\Dd^2\D^{\a_s}H^{(2)}_{\a(s)\ad(s)}\cr
}
\ee

\be
\eqalign{
{\cal G}^{(2)}_{\a(s)\ad(s)}=&+\frac{2}{s!}\bar{\D}_{(\dot{\alpha}_s}\D^2\Psi_{\alpha(s)\dot{\alpha}(s-1))}-\frac{2}{s!}\D_{(\alpha_s}\bar{\D}^2\bar{\Psi}_{\alpha(s-1))\dot{\alpha}(s)}\label{G2}\cr
&+2\D^{\gamma}\bar{\D}^2\D_{\gamma}H^{(2)}_{\alpha(s)\dot{\alpha}(s)}\cr
}
\ee

From the above we find:
\be
\eqalign{
&\pa_{(\a_{2s}}{}^{\ad_1}\dots\pa_{\a_{s+2}}{}^{\ad_{s-1}}\D_{\a_{s+1}}\Dd^2{\cal T}_{\alpha(s))\dot{\alpha}(s-1)}=\label{A2}\cr
&~~~~~~~~~~~~=-2\frac{s+1}{s}\pa_{(\a_{2s}}{}^{\ad_1}\dots\pa_{\a_{s+2}}{}^{\ad_{s-1}}\D_{\a_{s+1}}\Dd^2\D^2\Psi_{\a(s))\ad(s-1)}\cr
&~~~~~~~~~~~~~~+2i\pa_{(\a_{2s}}{}^{\ad_1}\dots\pa_{\a_{s+2}}{}^{\ad_{s-1}}\pa_{\a_{s+1}}{}^{\ad_{s}}\D_{\a_{s}}\Dd^2\bar{\Psi}_{\a(s-1))\ad(s)}\cr
&~~~~~~~~~~~~~~+2i\pa_{(\a_{2s}}{}^{\ad_1}\dots\pa_{\a_{s+2}}{}^{\ad_{s-1}}\D_{\a_{s+1}}\Dd^2\D^{\b}\pa_{\b}{}^{\ad_s}H^{(2)}_{\a(s))\ad(s)}\cr
}
\ee

\be
\eqalign{
&\pa_{(\a_{2s}}{}^{\ad_1}\dots\pa_{\a_{s+1}}{}^{\ad_{s}}\D_{\a_{s}}\bar{\cal T}_{\alpha(s-1))\dot{\alpha}(s)}=\label{B2}\cr
&~~~~~~~~=-2\frac{s+1}{s}\pa_{(\a_{2s}}{}^{\ad_1}\dots\pa_{\a_{s+1}}{}^{\ad_{s}}\D_{\a_{s}}\Dd^2\bar{\Psi}_{\a(s-1))\ad(s)}\cr
&~~~~~~~~~~~-2i\pa_{(\a_{2s}}{}^{\ad_1}\dots\pa_{\a_{s+2}}{}^{\ad_{s-1}}\D_{\a_{s}}\Dd^2\D^2\Psi_{\a(s))\ad(s-1)}\cr
&~~~~~~~~~~~-2\pa_{(\a_{2s}}{}^{\ad_1}\dots\pa_{\a_{s+1}}{}^{\ad_{s}}\D_{\a_{s}}\Dd^2\D^{\b}H^{(2)}_{\b\a(s-1))\ad(s)}\cr
}
\ee
and these two equations \eqref{A2} and \eqref{B2}, combined give:
\be
\eqalign{ {~~~~}
&\pa_{(\a_{2s}}{}^{\ad_1}\dots\pa_{\a_{s+2}}{}^{\ad_{s-1}}\D_{\a_{s+1}}\Dd^2{\cal T}_{\alpha(s))\dot{\alpha}(s-1)}-i\pa_{(\a_{2s}}{}^{\ad_1}\dots\pa_{\a_{s+1}}{}^{\ad_{s}}\D_{\a_{s}}\bar{\cal T}_{\alpha(s-1))\dot{\alpha}(s)}=\label{C2}\cr
&~~~~~~~~~~-2\frac{2s+1}{s}\pa_{(\a_{2s}}{}^{\ad_1}\dots\pa_{\a_{s+2}}{}^{\ad_{s-1}}\D_{\a_{s}}\Dd^2\D^2\Psi_{\a(s))\ad(s-1)}\cr
&~~~~~~~~~~+2i\frac{2s+1}{s}\pa_{(\a_{2s}}{}^{\ad_1}\dots\pa_{\a_{s+1}}{}^{\ad_{s}}\D_{\a_{s}}\Dd^2\bar{\Psi}_{\a(s-1))\ad(s)}\cr
&~~~~~~~~~~+2i\frac{2}{(2s)!}\D_{(\a_{2s}}\Dd^2X_{\a(2s-1))}
}
\ee
where
\be
\eqalign{
X_{\a(2s-1)}=&+s\pa_{(\a_{2s-1}}{}^{\ad_1}\dots\pa_{\a_{s+1}}{}^{\ad_{s-1}}\D^{\b}\pa_{\b}{}^{\ad_s}H_{\a(s))\ad(s)}\cr
&+s\pa_{(\a_{2s-1}}{}^{\ad_1}\dots\pa_{\a_{s}}{}^{\ad_{s}}\D^{\b}H_{\b\a(s-1))\ad(s)}\cr
}
\ee

From \eqref{G2} we see:
\be
\eqalign{
&\pa_{(\a_{2s}}{}^{\ad_1}\dots\pa_{\a_{s+1}}{}^{\ad_{s}}{\cal G}^{(2)}_{\a(s))\ad(s)}=\label{D2}\cr
&~~~~~~~~~~~~~~~~~=-2i\pa_{(\a_{2s}}{}^{\ad_1}\dots\pa_{\a_{s+2}}{}^{\ad_{s-1}}\D_{\a_{s+1}}\Dd^2\D^2\Psi_{\alpha(s))\dot{\alpha}(s-1)}\cr
&~~~~~~~~~~~~~~~~~~~~-2\pa_{(\a_{2s}}{}^{\ad_1}\dots\pa_{\a_{s+1}}{}^{\ad_{s}}\D_{\alpha_s}\bar{\D}^2\bar{\Psi}_{\alpha(s-1))\dot{\alpha}(s)}\cr
&~~~~~~~~~~~~~~~~~~~~+\frac{2}{2s+1}\D^{\gamma}\bar{\D}^2\D_{(\gamma}\pa_{\a_{2s}}{}^{\ad_1}\dots\pa_{\a_{s+1}}{}^{\ad_{s}}H^{(2)}_{\alpha(s))\dot{\alpha}(s)}\cr
&~~~~~~~~~~~~~~~~~~~~-2\frac{2s}{(2s+1)!}\D_{(\a_{2s}}\Dd^2X_{\a(2s-1))}
}
\ee
due to the identity:
\be
\eqalign{
\D_{\gamma}\pa_{(\a_{2s}}{}^{\ad_1}\dots\pa_{\a_{s+1}}{}^{\ad_{s}}H^{(2)}_{\alpha(s))\dot{\alpha}(s)}=&
\frac{1}{2s+1}\D_{(\gamma}\pa_{\a_{2s}}{}^{\ad_1}\dots\pa_{\a_{s+1}}{}^{\ad_{s}}H^{(2)}_{\alpha(s))\dot{\alpha}(s)}\cr
&-\frac{2s}{(2s+1)!}C_{\gamma(\a_{2s}}X_{\a(2s-1))}
}
\ee
At this point we can define another chiral field strength superfield
\be
\eqalign{
{\cal W}_{\a(2s+1)}=\bar{\D}^2\D_{(\a_{2s+1}}\pa_{\a_{2s}}{}^{\ad_1}\dots\pa_{\a_{s+1}}{}^{\ad_{s}}H^{(2)}_{\alpha(s))\dot{\alpha}(s)}  ~~~.
}
\ee
Finally equations \eqref{C2}, \eqref{D2} and the adove definition when combined, give the following Bianchi identity

\be
\eqalign{
\D^{\a_{2s+1}}{\cal W}_{\a(2s+1)}=&\frac{2s+1}{2}\pa_{(\a_{2s}}{}^{\ad_1}\dots\pa_{\a_{s+1}}{}^{\ad_{s}}{\cal G}^{(2)}_{\a(s))\ad(s)}\cr
&-i\frac{s}{2}\pa_{(\a_{2s}}{}^{\ad_1}\dots\pa_{\a_{s+2}}{}^{\ad_{s-1}}\D_{\a_{s+1}}\Dd^2{\cal T}_{\alpha(s))\dot{\alpha}(s-1)}\cr
&+\frac{s}{2}\pa_{(\a_{2s}}{}^{\ad_1}\dots\pa_{\a_{s+1}}{}^{\ad_{s}}\D_{\a_{s}}\bar{\cal T}_{\alpha(s-1))\dot{\alpha}(s)}\cr
}
\ee

and by definition
\be
\eqalign{
\Dd_{\ad}{\cal W}_{\a(2s+1)}=0
}
\ee

On-Shell the superfield strengths vanish (equations of motion)
\bse
\bea
&&{\cal T}_{\a(s)\ad(s-1)}=0\\
&&\bar{\cal T}_{\a(s-1)\ad(s)}=0\\
&&{\cal G}^{(2)}_{\a(s)\ad(s)}=0  ~~.
\eea
\ese

Hence we get
\bse
\bea
&\D^{\a_{2s+1}}{\cal W}_{\a(2s+1)}=0 ~,~ &\Dd^{\ad_{2s+1}}\bar{\cal W}_{\ad(2s+1)}=0\\
&\Dd_{\ad}{\cal W}_{\a(2s+1)}=0 ~,~ &\D_{\a}\bar{\cal W}_{\ad(2s+1)}=0 ~~~.
\eea
\ese

This system describes a half odd superspin $Y=s+1/2$ massless supermultiplet


\section{Conclusion}

~~~In this work, we have presented a unified treatment of the work of Kuzenko et.\ al. Along the
way, we have provided the first (to our knowledge) derivation of the explicit forms of the field
strength superfields in terms of prepotentials and associated Bianchi identities.   Our investigation
also sets the stage for the study of possible alternative off-shell formulations of higher superspin
superfield theories.  The departure for this is to consider a generalization of the gauge transformation
in (4) to include linear combinations of all the terms that appear in (3).  This more general study
will be undertaken in a later effort.


{\bf Acknowledgements}:

We want to thank G. Tartaglino-Mazzuccelli for discussion.  SJG wishes to acknowledge
S.\ Kuzenko for conversations and as well I.\ McArthur along with the University of Western
Australian School of Physics and UWA Institute for Advanced Study for hospitality.
This research was supported in part by the endowment of the John S.~Toll Professorship,
the University of Maryland Center for String \& Particle Theory, and National Science Foundation 
Grant PHY-0354401. 

\end{document}